# Trapping light in air with membrane metasurfaces for vibrational strong coupling

Wihan Adi[1,‡], Samir Rosas[1,‡], Aidana Beisenova[1], Shovasis Kumar Biswas[2], Hongyan Mei[2], David A. Czaplewski[3], Filiz Yesilkoy[1,*]

[1] Department of Biomedical Engineering, University of Wisconsin-Madison Madison, WI 53706, USA
[2] Department of Electrical and Computer Engineering, University of Wisconsin-Madison Madison, WI 53706, USA
[3] Center for Nanoscale Materials, Argonne National Laboratory, Lemont, IL, 60439, USA

*Email: filiz.yesilkoy@wisc.edu
‡ Authors contributed equally

**Abstract**

Optical metasurfaces can manipulate electromagnetic waves in unprecedented ways at ultra-thin engineered interfaces. Specifically, in the mid-infrared (mid-IR) region, metasurfaces have enabled numerous biochemical sensing, spectroscopy, and vibrational strong coupling (VSC) applications via enhanced light-matter interactions in resonant cavities. However, mid-IR metasurfaces are usually fabricated on solid supporting substrates, which degrade resonance quality factors (Q) and hinder efficient sample access to the near-field electromagnetic hotspots. Besides, typical IR-transparent substrate materials with low refractive indices, such as $CaF_2$, NaCl, KBr, and ZnSe, are usually either water-soluble, expensive, or not compatible with low-cost mass manufacturing processes. Here, we present novel free-standing Si-membrane mid-IR metasurfaces with strong light-trapping capabilities in accessible air voids. We employ the Brillouin zone folding technique to excite tunable, high-Q quasi-bound states in the continuum (q-BIC) resonances with our highest measured Q-factor of 722. Leveraging the strong field localizations in accessible air cavities, we demonstrate VSC with multiple quantities of PMMA molecules and the q-BIC modes at various detuning frequencies. Our new approach of fabricating mid-IR metasurfaces into semiconductor membranes enables scalable manufacturing of mid-IR photonic devices and provides exciting opportunities for quantum-coherent light-matter interactions, biochemical sensing, and polaritonic chemistry.

**Introduction**

Optical metasurfaces are ultra-thin engineered interfaces characterized by sub-wavelength periodic structures, capable of exhibiting unique photonic properties distinct from those of the bulk materials from which they are derived. Recent developments in the photonics field have shown that optical metasurfaces enable unprecedented functionalities via the manipulation of electromagnetic waves at various degrees of freedom[1–4]. By leveraging diverse resonance mechanisms[5–7], these engineered interfaces can enable on-chip wavefront shaping[8], phase, direction[3,9], and polarization[10] control. In addition to their free-space wave modulation capabilities, resonant metasurfaces can spatially and temporally confine light, generating powerful photonic cavities in the vicinity of resonators. Specifically, dielectric metasurfaces supporting quasi-bound states in the continuum (q-BIC) resonances have recently become a new paradigm for studying light-matter interactions due to their high quality-factor (Q) cavities[11,12], where intense near-field hotspots form. From ultraviolet[13] to terahertz[14] spectral ranges, q-BIC metasurfaces have enabled exciting photonic applications, including biochemical sensing[15–17], lasing[18], and quantum optics[19,20]. In particular, in the mid-infrared (mid-IR) spectral region, where many functional molecules have their vibrational states, q-BIC metasurfaces can enhance the analytical performance of infrared absorption spectroscopy[21], enabling various biochemical sensing applications[22]. Moreover, q-BIC metasurfaces are excellent candidates for achieving light-mediated control over molecules' innate quantum states via vibrational strong coupling (VSC) between the high-Q mid-IR cavity resonances and the molecule's ground vibrational states[23–25]. However, to unleash the full potential of mid-IR metasurfaces as effective biochemical sensors and chemical process modulators, their fundamental material and fabrication constraints need to be addressed.

Typical q-BIC metasurfaces are fabricated by first depositing a layer of high-index and low-loss material (e.g., Si, Ge) on a solid substrate and patterning resonator arrays into the deposited layer. In such all-dielectric metasurfaces, a key requirement to generate high-Q optical resonances is to maximize the refractive index contrast between the high-index resonators and their surrounding media[26–30]. Due to their low absorption losses, Si and Ge are widely used materials in mid-IR metasurface fabrication. However, Si and Ge cannot be used simultaneously for resonator and supporting substrate material because this would diminish the index contrast and hinder the metasurface photonic functionalities. Therefore, materials with low absorption loss that are transparent over an IR window, such as $CaF_2$, NaCl, KBr, and ZnSe, are primarily used as supporting substrates in mid-IR metasurfaces. Yet, these materials are either water-soluble, too brittle to be reliably used in large-scale manufacturing processes, or expensive. The lack of CMOS-compatible, low-loss IR substrates, other than Si and Ge, presents a bottleneck for high throughput fabrication of mid-IR metasurfaces. Additionally, it is common for the resonance modes to leak into the substrates, obstructing access to electromagnetic hotspots, which is critical for efficient near-field light-matter interactions[27]. Previously, free-floating thin membranes were proposed as supporting layers for mid-IR metasurfaces[31]. While this approach eliminates the need for bulk IR substrates, mechanically robust membrane materials, such as $Al_2O_3$, and $Si_3N_4$[32,33], are lossy in the mid-IR, and despite being thin, they introduce undesired asymmetry in the dielectric environment surrounding the resonators. Thus, substrates that are used as metasurface carriers perturb the critical photonic functionalities of the dielectric metasurfaces while obstructing their industrial-scale manufacturability and technology transfer, which calls for a novel solution.

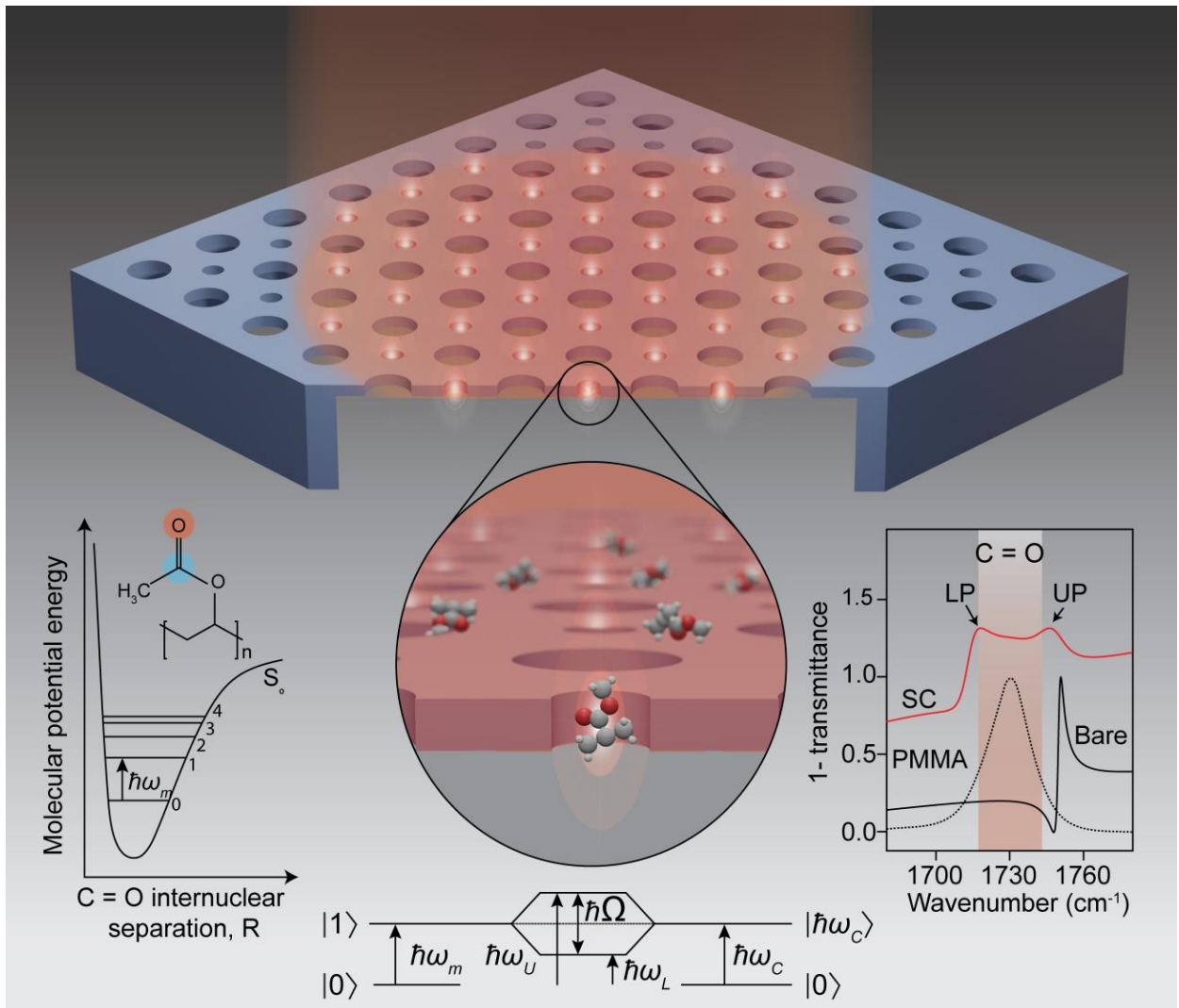

**Fig. 1: Artistic rendering of qBIC-supporting, free-standing Si-membrane metasurfaces enabling vibrational strong coupling; objects in the image do not represent their realistic dimensions.** The Brillouin zone folding (BZF) method is used to excite q-BIC modes and efficiently trap light into air voids. Upon coupling the q-BIC air mode to the C=O vibrational mode of PMMA molecules, vibrational polaritons are generated, revealing a coherent energy exchange between the cavity and the molecule. Schematics depict the all-Si metasurface with mid-IR light trapping capabilities and the plot on the right shows the simulation of vibrational strong coupling between the molecular vibrational states and the qBIC cavity resonance mode.

Here, we present a novel concept of free-standing qBIC-supporting metasurfaces that can efficiently trap mid-IR light in air cavities for VSC (Fig. 1). Our approach to fabricating low-loss metasurfaces is by perforating carefully designed periodic hole arrays into a free-standing high-index Si-membrane. This approach circumvents the necessity of introducing additional materials into the fabrication process and leverages the well optimized fabrication processes of Si. As illustrated in Fig. 1, this low-loss, tunable optical cavity can enable coherent energy exchange between the cavity and molecules, forming vibrational polaritons in the frequency domain. By periodically perturbing the size of the hole radii (Fig. 2a, b), we engineer the size of the Brillouin zone in the *k*-space and modify the light dispersion band folding scheme. In this unique design,

Brillouin zone folding (BZF) introduces a leak into the trapped guided modes (GMs), converting them to q-BIC resonances, also called BZF-induced BIC modes, previously explored in the near-IR and THz domains[34,35]. Contrary to typical photonic guided modes supported by photonic crystals, BZF-induced qBIC modes can efficiently trap light ($|E|/|E_0|$~38) in air voids (Fig. 2i and S4), generating strong cavities suited for near-field light-matter coupling. The design is etched into a thin (1 µm) Si-membrane (Fig. 2c-f), which eliminates the need for IR-transparent carrier substrates (Fig. 2g). Furthermore, typical in qBIC modes, we show control over radiation channels via a radius perturbation parameter *Δr* (Fig. 2h), and our highest measured Q-factor of 722 (Fig. 3d). Moreover, we numerically and optically characterize the fabricated metasurfaces by identifying their resonance types and mode properties evaluating their suitability for enhancing near-field light-matter interactions. Finally, we demonstrate the VSC between the qBIC resonances and the C=O stretching mode of the polymethyl methacrylate (PMMA) molecules at 1730 cm$^{-1}$ by identifying the polariton formation and measuring the signature Rabi mode splitting over a range of detuning frequencies $\delta = \omega_{\text{PMMA}} - \omega_{\text{qBIC}}$. Our experimental findings closely correlate with the numerically calculated spectral anti-crossing pattern and reveal the collective VSC strength as a function of material quantity in the qBIC cavity. Altogether, we introduce a new class of free-standing Si-membrane metasurfaces that are completely CMOS manufacturing compatible and can impact light-matter interactions from weak to strong coupling in powerful open cavities in the mid-IR.

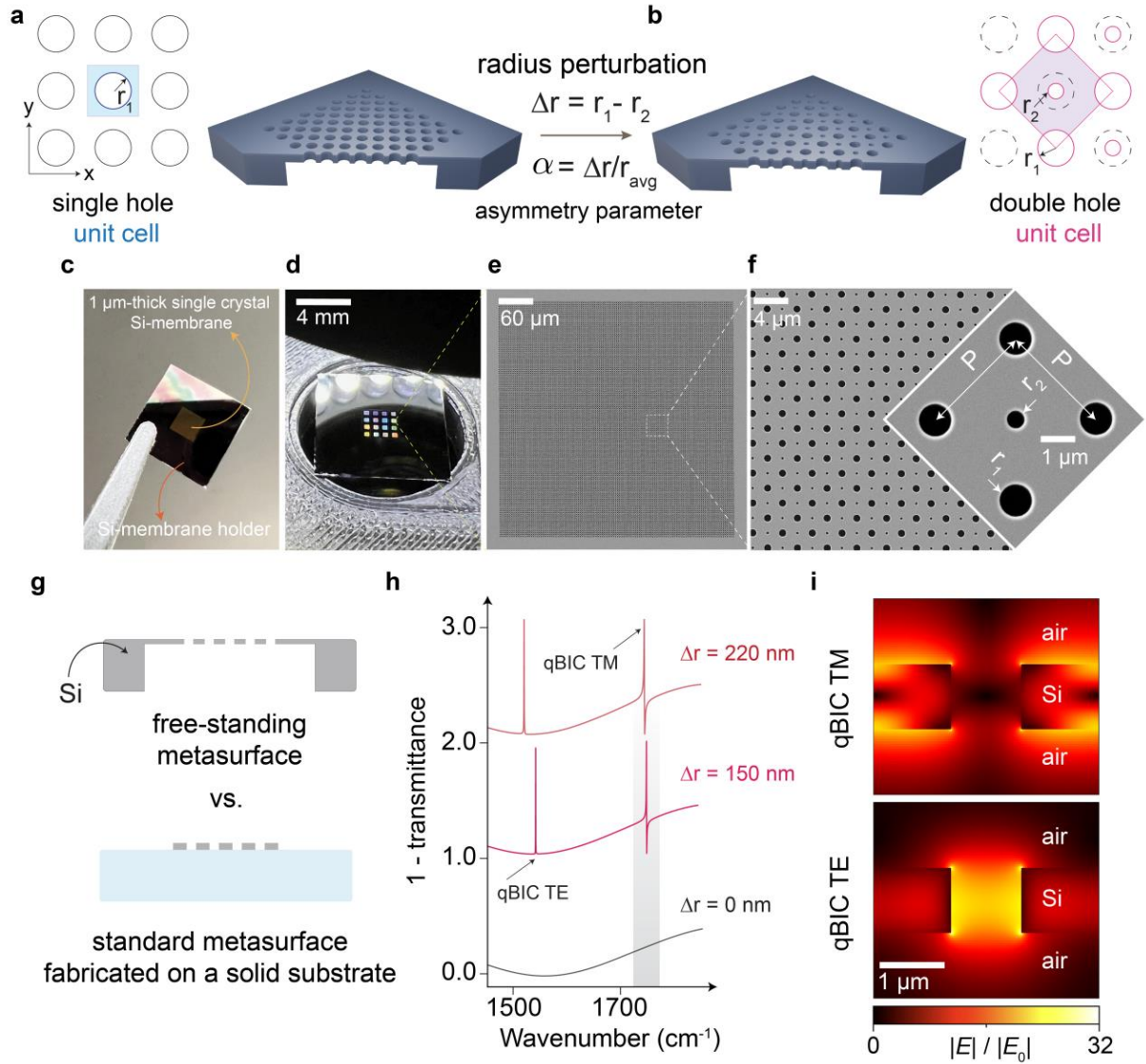

**Fig. 2: BZF-qBIC resonances supported by free-standing Si-membrane metasurfaces. a, b,** Schematics of periodic radius perturbation in a 2-D square-lattice hole array to create a qBIC-supporting double-hole design in Si-membranes. **c,** Photograph of an unpatterned free-standing Si-membrane chip with a membrane window of 2.8 mm x 2.8 mm. **d,** Photograph of a 4x4 array of BZF-qBIC-supporting metasurfaces, where distinct colors captured by the camera in the visible correspond to variations in the geometric parameters of the double-hole unit cell. **e, f,** Scanning electron microscopy (SEM) images of the metasurface and a double-hole unit cell with its critical geometric parameters. **g,** Schematic to illustrate the comparison between metasurfaces with and without substrate. **h,** Simulated spectra showing the emerging BZF-qBIC modes when the symmetry is broken, i.e., $\Delta r \neq 0$; shaded gray area is the full-width-half-maximum (FWHM) of PMMA C=O vibrational bond that is 26.8 cm$^{-1}$ wide. **i,** Simulated electric field enhancement maps for supported qBIC TE and qBIC TM modes with 1 μm scale bar.

## Results

### Metasurface structure and fabrication

To fabricate the substrate-less all-Si metasurfaces, we used free-standing single-crystal <100> Si-membranes (thickness=1 µm) with 2.8 mm X 2.8 mm dimensions (Fig. 2c). The metasurface designs were patterned on the membranes using electron beam lithography, and then deep reactive ion etching was used to transfer the patterns by etching through the membranes and forming air holes (Fig. S1a). This is contrary to conventional dielectric metasurfaces that are composed of an array of high-index subwavelength structures fabricated on solid substrates. Fig. S1b shows a typical fabrication sequence of standard metasurface devices. Herein, we leverage an inverted approach where subwavelength resonator arrays of low-index air holes are buried in a free-standing high-index dielectric membrane; thus, we eliminate the need for bulk supporting substrates. Fig. 2d shows a typical fabricated 4-by-4 metasurface array, where each metasurface dimension is 300 µm × 300 µm. Scanning electron microscopy (SEM) images of the fabricated metasurfaces with a double-hole design are shown in Fig. 2e, and 2f, where the inset shows a repeating pattern (metaunit) with its critical geometric design parameters that are used to tune the optical resonance properties.

### Brillouin zone folding (BZF) induced quasi-bound states in continuum (q-BIC) resonances

Our approach to engineering BZF-qBIC supporting metasurfaces is introducing periodic perturbations to a photonic crystal slab where the metaunit is a square with a hole at its center, which we call the single-hole design (Fig. 2a). To bring the photonic crystal guided modes (GMs) to the mid-IR region, our single-hole design features hole radii of $r$=0.9 µm with periods $P$=2.46 µm. Then, we reduce the radii of every other hole in both x and y directions by $\Delta r$, which changes the unit cell and enlarges its period by a factor of $\sqrt{2}$ to $P$=3.48 µm (Fig. 2b). We call this radius perturbed metasurface design the double-hole. Fig S2a and S2c show the simulated TE and TM transmission spectra of the single-hole and double-hole metasurfaces ($\Delta r$=0.36 µm) that are excited with normal incident light (Γ-point). At the Γ-point, the double-hole structure's dispersion relation has qBIC and GM modes, whereas the single-hole structure supports only the GM modes. The TE and TM modes are excited separately in the simulation by setting the appropriate symmetry conditions with respect to the mirror plane parallel to the membrane, even for TE and odd for TM[36]. Further simulation details can be found in the methods section. In the double-hole design, periodic radius perturbation shifts the Brillouin zone from X to X', folding all the fundamental photonic crystal TE and TM modes (Fig. 3a, S2b, and S2d). The qBIC modes emerge due to the rearranged modes at lower $k$-values, enabling free-space excitation of the GMs originally located below the light cone. Notably, the radius perturbation in the double-hole design introduces radiative leakage channels and thus enables free-space access to the symmetry-protected BIC modes at the Γ-point (Fig. S2a and S2c). Thus, BZF-qBICs emerge as free-space-accessible resonances at the Γ-point, whose Q-factors are controlled by perturbation, $\Delta r$, in contrast to their counterparts in the single-hole design, which are inaccessible—hence their infinite lifetime[35]. The simulation results (Figs. S3a, S3b) and measured transmittance spectra (Fig. 3c) show the correlation between the BZF-qBIC modes' Q-factors and the $\Delta r$, radius perturbation.

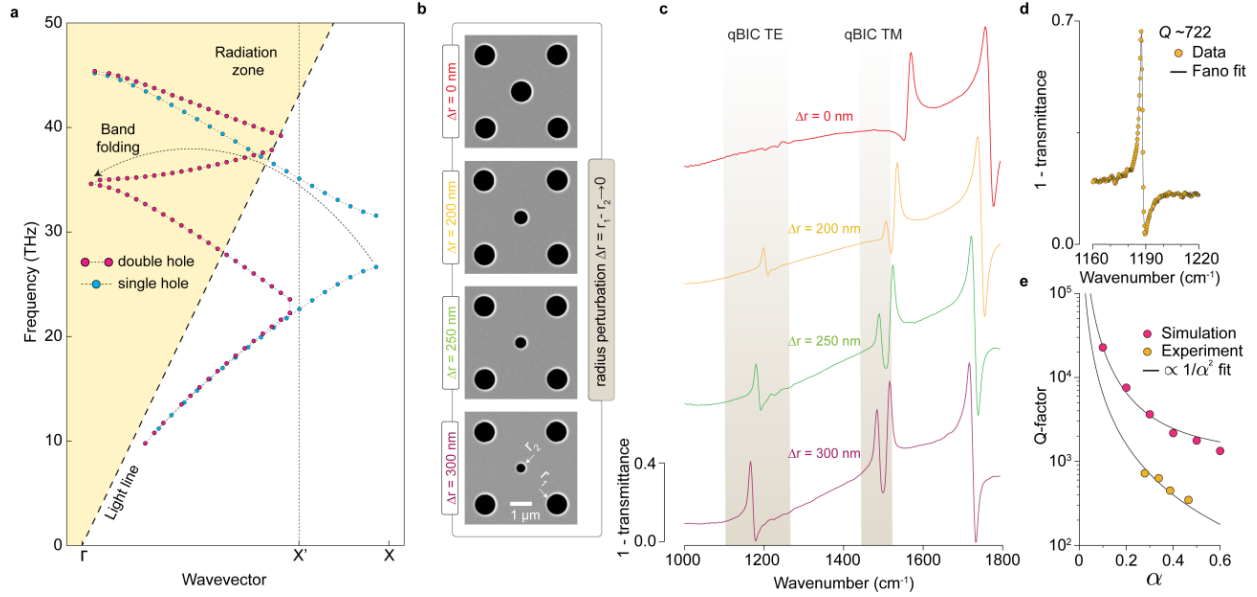

**Fig. 3: Free-space accessible BZF-qBIC modes supported by Si-membrane metasurfaces in the mid-IR. a,** Calculated band diagram of the TE modes supported by single (blue) and double-hole (red) designs. X is the edge of the Brillouin zone of the single-hole, and X' is that of the double-hole metasurface designs. **b, c,** SEM images of the fabricated metasurfaces' unit cells with various *Δr* values and their corresponding spectral responses demonstrating the resonance modes. The spectral region of qBIC TE and TM modes are shaded in brown; no qBIC modes are detected when *Δr*=0 nm, because of its high (theoretically infinite) Q-factor. **d,** The highest measured Q-factor of 722 for the TE-mode from the fabricated device of *Δr*=200 nm/ *α*=0.28. **e,** Evolution of Q-factor as a function of asymmetry parameter $\alpha = \Delta r / r_{avg}$ for simulated and experimental data.

We further analyzed the near-field characteristics of the resonances supported by the single-hole and double-hole metasurfaces to investigate their suitability for cavity-enhanced light-matter interaction studies. Fig. S4 a-d shows the electric (E) field enhancement profiles of qBIC TE, GM TE, qBIC TM, and GM TM modes across the x-z plane of the membrane, which extends in the x-y plane. Fig. S4e and S4g show the field enhancement profiles along the corresponding dashed lines indicated on the maps. Simulation results reveal that qBIC modes can enhance the E-field significantly better than the GMs (~ 5-fold, 38 vs. 7 for TE and 29 vs. 6 for TM). The E-field enhancement values are in a similar range to those reported using other qBIC metasurface designs (~10–40)[21,37,38], and much higher than a Fabry-Perot cavity (~3, Fig. S17), a commonly used photonic cavity design for strong coupling experiments. Notably, in our design, the electric field hotspots are predominantly formed on the exterior surface of the metasurface. Specifically, the qBIC TE mode can efficiently trap the E-field in the center of the smaller air holes, whereas qBIC TM modes concentrate theirs on the membrane surface. This is because the TE modes are even with respect to the mirror plane parallel to the membrane, and thus, its electric field direction is along the slab[36]. This is in contrast to the odd symmetry of TM modes, which makes its E-field direction perpendicular to the slab[36]. Furthermore, we calculated the bulk refractive index sensitivity ($S = \Delta\lambda/\Delta n$) of each mode (Fig. S4f and S4h). Eqn. 1 (from[39]) lays out the relation between resonance frequency shift *Δω* of a mode for a given permittivity perturbation *Δε*. Resonance frequency is denoted as *ω*, resonance frequency shift *Δω*, electric field *E*, and permittivity of the perturbation *Δε*. Because qBIC modes can effectively concentrate light in air cavities, their refractive index sensitivities (~1500 nm/RIU for qBIC-TM and ~1400 nm/RIU for qBIC-TE) are significantly higher than the GMs and the previously reported nanophotonic structures[15,40].

$$\Delta\omega = -\frac{\omega}{2}\frac{\int d^3\boldsymbol{r}\,\Delta\varepsilon(\boldsymbol{r})|\boldsymbol{E}(\boldsymbol{r})|^2}{\int d^3\boldsymbol{r}\,\varepsilon(\boldsymbol{r})|\boldsymbol{E}(\boldsymbol{r})|^2} \qquad (1)$$

Moreover, we calculated mode volumes of each BZF-qBIC resonances per unit cell as 0.42 $(\lambda/n)^3$ for qBIC-TE, and 0.96 $(\lambda/n)^3$ for qBIC-TM, which are significantly lower than frequently used photonic cavities, such as Fabry-Perot (~ 5 $(\lambda/n)^3$) and whispering gallery modes[41] (~6 $(\lambda/n)^3$).

Subsequently, we fabricated the double-hole metasurfaces with varying radius perturbations. Fig. 3b shows the SEM images of the metaunits of four different designs with $\Delta r$ =0, 200, 250, and 300 nm, which correspond to the asymmetry parameters, $\alpha=\Delta r/r_{avg}$, of 0, 0.24, 0.31, and 0.4, respectively. To optically characterize the fabricated metasurfaces, we primarily used a tunable quantum cascade laser (QCL)-based microscope with a minimum 2 cm$^{-1}$ spectral tunability limit. Thus, we chose the $\Delta r$ values large enough (~ 300 nm) to be able to measure the resonances on our main optical setup. Fig. 3c shows the measured spectral responses with varying radius perturbations. In accordance with the simulation results (Fig. S3), the TE and TM qBIC mode amplitudes decrease as $\Delta r$ approaches zero until they finally disappear at $\Delta r$=0 nm. Since our illumination can excite both TE and TM modes, each measured spectrum shows all four modes (TE, TM of both GMs and qBICs). While the GMs' linewidths are barely affected by the change of $\Delta r$, qBIC resonances' Q-factor increased and amplitude decreased with decreasing $\Delta r$. This is caused by the spectral resolution limit of the optical setup failing to capture the ultra-sharp resonance peaks and the suppressed radiation channels as the Q-factor increases with decreasing $\Delta r$[35].

For a more precise Q-factor characterization of the sharp BZF-qBIC modes, we used a Fourier Transform Infrared Spectrometer (FTIR) attached to a mid-IR microscope. We used low-NA refractive objectives for close-to-normal transmission measurements. We measured a maximum Q-factor of 722 for the qBIC TE mode (Fig 3d) and 463 for the qBIC-TM mode. We also plotted the FTIR-measured and simulation-estimated Q-factor values of the qBIC TE mode and observed a quadratic relationship between Q and the asymmetry parameter $\alpha$ (Fig. 3e), which is a signature trend in numerous metasurfaces supporting qBIC resonances[42].

Moreover, to demonstrate the spectral tunability of the qBIC modes, we adjusted the metaunit design parameters by scaling parameter $s$ (Fig. 4a, inset). Fig. 4a and 4b show the simulated and measured qBIC TM resonances, respectively, with peaks tuned to different wavenumbers. Here, we specifically targeted the absorption peak associated with the C=O vibrational bond of the PMMA molecule, which peaks around 1730 cm$^{-1}$ with a full-width half maximum (FWHM) of 26.8 cm$^{-1}$ (Fig. S5, also shaded in gray in Fig. 4a and 4b). Expecting a red shift in the resonance peak upon introducing the PMMA molecules into the cavities due to the change of the surrounding refractive index, we fabricated a set of metasurfaces with resonance peaks slightly higher in wavenumbers (shorter wavelengths) than 1730 cm$^{-1}$. The average FWHM of the fabricated qBIC TM resonances shown in Fig. 4b is ~ 13 cm$^{-1}$, i.e., the average loss is 6.5 cm$^{-1}$ (Fig. S16), which is a compatible spectral width with the spectral resolution of our QCL-based microscope (2cm$^{-1}$).

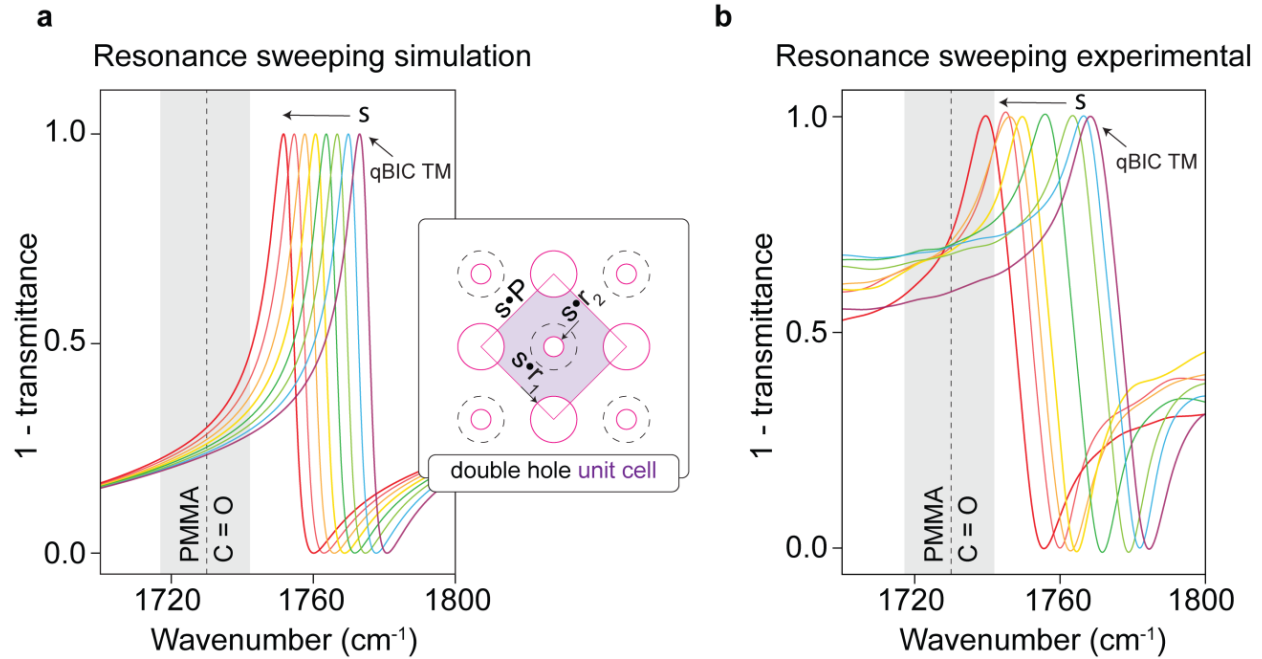

**Fig. 4: Resonance sweeping of the BZF-qBIC metasurfaces. a, b,** The spectral positions of the qBIC TM resonance modes are tuned by varying the geometric metaunit parameters in double-hole design using scaling parameter *s* (inset, a). Numerically simulated (a) and measured (b) spectra of the TM modes tuned to a higher wavenumber region than the C=O absorption band to accommodate for the expected redshift upon PMMA deposition into the cavities.

**Vibrational strong coupling (VSC) with BZF-qBIC cavities**

To demonstrate the VSC capabilities of the double-hole metasurfaces, we studied the coupling strength between the vibrational transition associated with the C=O bond of PMMA molecules and the qBIC TM resonances. We first coated the metasurface array whose resonance peaks (shown in Fig. 4) were tuned to the right of 1730 cm$^{-1}$ (Fig. 5a) with a PMMA layer ($t$=150 nm). Fig. 5a and S8a show the measured and simulation-derived absorbance spectra, respectively, as the cavity resonance sweeps through the absorption peak. The spectra shown in Fig. 5a were normalized to their maximum, and their corresponding raw spectra are shown in Fig. S9. Due to the refractive index increase with PMMA coating, the cavity resonances shift to lower wavenumbers, which can be observed via the uncoated metasurfaces' resonance peaks, plotted in Fig. 5a in dashed lines for reference. When the cavity resonance starts overlapping with the molecules' vibrational mode, a polariton pair forms. The separation between the lower and upper polariton is defined by the Rabi constant in Eqn. 2 (from[43]), which correlates to the coherent energy exchange between the cavity and molecular states:

$$\Omega = \sqrt{4g^2 + (\delta - i(\gamma_{PMMA} - \gamma_{qBIC}))^2} \qquad (2)$$

where the $\delta = \omega_{PMMA} - \omega_{qBIC}$ is the detuning parameter, $\gamma_{PMMA}$ and $\gamma_{qBIC}$ are PMMA and cavity losses, respectively, and $g$ is the coupling strength, defined as $g = d \cdot E = \sqrt{(\hbar \cdot \omega)/(2 \cdot \varepsilon \cdot \varepsilon_0 V)}$, where $d$ is the transition dipole moment, $E$ is the cavity's vacuum electric field, $\varepsilon$ and $\varepsilon_0$ are material and vacuum permittivity, respectively, and $V$ is the mode volume[43]. The Rabi splitting reaches a minimum at zero detuning frequency $\delta = \omega_{PMMA} - \omega_{qBIC} = 0$. This trend matches with our experimental and simulation results (Fig. 5b and S8b) and other work[25]. Moreover, we present the anti-crossing behavior of the cavity resonance peak position as it sweeps the molecular vibrational transition band (Fig. 5c and S8c). This is further illustrated by simulation using a higher density of *s* parameters shown in Fig. 5d. At zero detuning, we observed a Rabi splitting of 20 cm$^{-1}$, which is right around the onset of strong coupling criteria[43] of $\Omega > \gamma_{PMMA} + \gamma_{qBIC}$ = (13.4 + 6.5=19.9) cm$^{-1}$. When not overlapping with the PMMA resonance, the cavity resonance only shifts its peak wavenumber without much change in its loss (Fig. S14), and thus, the loss of the bare resonance can be used to approximate the cavity loss once coupled to PMMA C=O vibrational bond. Our measurements with PMMA layers >150 nm yielded Rabi splitting well above the strong coupling threshold of 19.9 cm$^{-1}$ (see Fig. 6b).

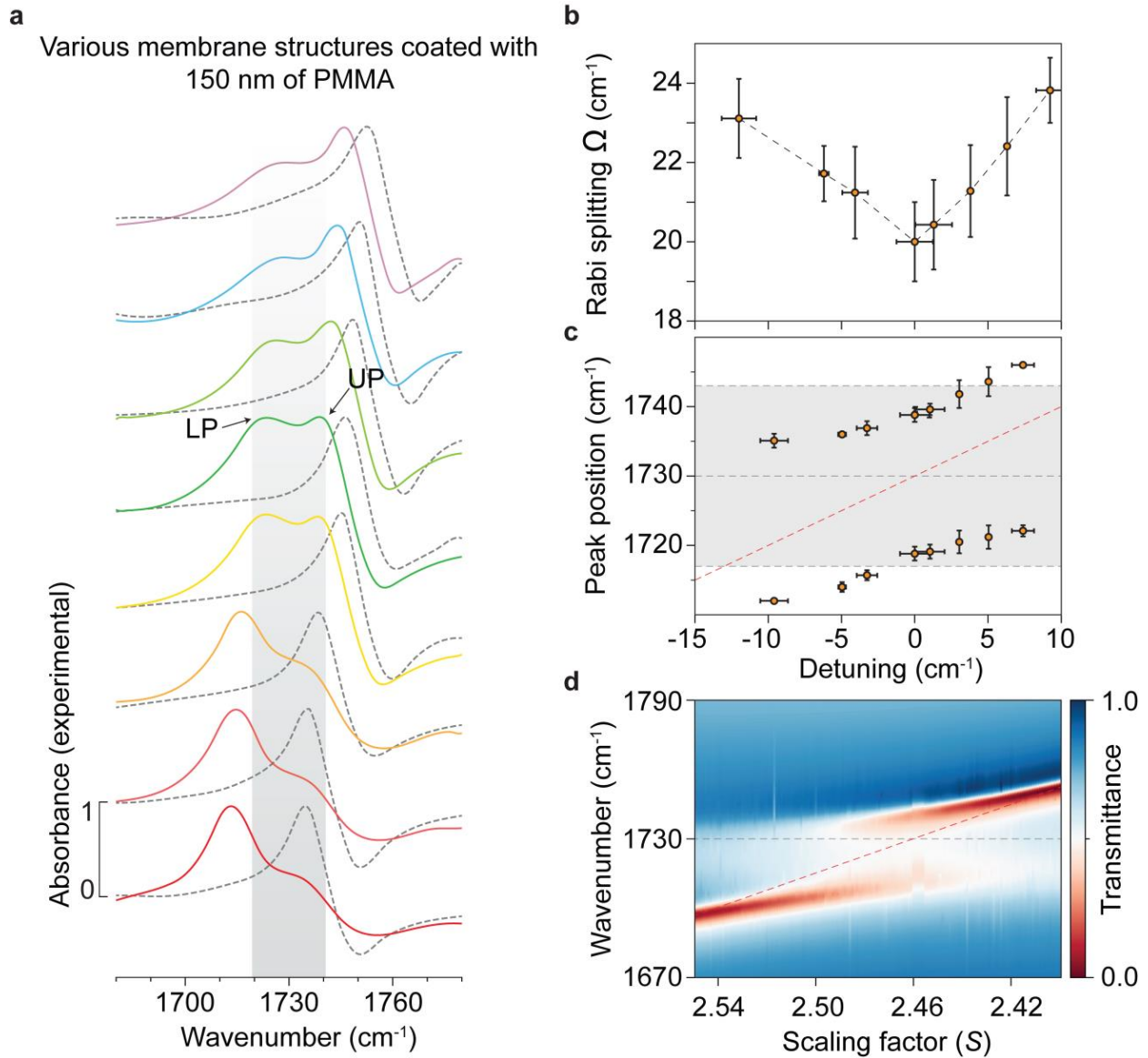

**Fig. 5: Rabi splitting as a function of frequency detuning between the PMMA vibrational mode and the metasurface resonances. a,** Spectra showing the VSC between the PMMA molecules' C=O vibrational transition (shaded gray) and the qBIC TM modes, which are spectrally tuned to various wavenumbers. Spectra collected before coating the metasurface with PMMA (dashed gray lines) and after coating (solid-colored lines) are shown. Experimentally measured spectral responses show polariton formation and Rabi splitting at varying values due to resonance detuning as the qBIC mode sweeps the C=O absorption band. Vertical offsets were applied to improve clarity. **b,** The experimentally measured Rabi splitting values plotted as a function of resonance detuning show the zero-detuning wavenumber at the minimum Rabi splitting value. **c, d,** The anti-crossing behavior, a signature of VSC, was observed both in experimental measurements (c) and simulation results (d). The gray dashed line represents PMMA C=O bond's absorbance resonance and the pink dashed line represents unperturbed qBIC TM resonances, respectively.

Furthermore, to investigate the effect of molecule quantity in the qBIC cavities, we studied VSC as a function of PMMA layer thickness $t$ coated on the metasurfaces. In this experiment, we worked with the qBIC TM mode shown in Fig. 4. because its field enhancement is primarily located on the top and bottom of the metasurface, decaying along the z-axis with a decay length of 650 nm (Fig. 6c and Fig. S4). This allowed us to control the number of interacting PMMA molecules with the cavity mode by varying the PMMA layer thickness *t* via spin-coating speed parameter. All PMMA thicknesses presented in Fig. 6a are within the field decay length, thus correlate with the number of molecules in the cavity. As the number of molecules ($N$) simultaneously coupled to a cavity increases, the coupling strength is boosted by a factor of $\sqrt{N}$ (from[44]), which is called collective strong coupling. We identified that the measured Rabi splitting is proportional to the square root of PMMA thickness (Fig. 6b), corroborating the collective VSC mechanism. For PMMA thicknesses above 260 nm, a central peak corresponding to the PMMA C=O band between the polaritons appears in the absorbance spectrum (Fig. S18 solid lines). We assumed that this is from non-polariton-forming PMMA molecules located further away from the membrane surface, thus experiencing a weaker electric field. To enhance the clarity of the polariton formation, we approximated the contribution from these non-polariton forming molecules to absorption from PMMA on the unpatterned region of the Si-membrane (dotted lines in Fig. S18) and subtracted them from the coupled systems' spectra, obtaining the signals shown in Fig. 6a. Due to the resolution limitations of our measurement setup, we could not resolve the polaritons for PMMA thicknesses equal to and less than 65 nm. However, a resonance peak broadening is apparent in the resonance spectrum of the metasurface with the thin (*t*=65 nm) PMMA coating compared to its uncoated counterpart. This effect is what is usually measured in biological and chemical sensors using surface enhanced infrared absorption (SEIRA) spectroscopy, which is based on the weak light-matter coupling phenomenon.

We have also shown polariton formations with the qBIC-TE mode and various thicknesses of PMMA (Fig. S6), measuring a relatively constant Rabi splitting of ~22 cm$^{-1}$. This value is close to the strong coupling limit obtained through the sum of PMMA loss (13.4 cm$^{-1}$, Fig. S5, and literature[45,46]) and bare cavity loss (~6.5 cm$^{-1}$, Fig. S15), i.e., $\gamma_{PMMA} + \gamma_{qBIC}$ = (13.4 + 6.5=19.9) cm$^{-1}$; thus, the system is at the onset of strong coupling. Contrary to the qBIC-TM mode measurements, we did not observe a variation in Rabi splitting as a function of the deposited PMMA layer thickness. This is because spin-coating does not allow for a controlled deposition of PMMA into the holes. The qBIC-TE mode has its field enhancement mostly within the hole, and therefore, based on our simulation results, extra PMMA deposition on the top surface does not significantly contribute to the polariton formation, but it causes damping and broadening of the mode (Fig. S13a). Meanwhile, PMMA deposition onto the inner walls of the holes leads to a significant Rabi splitting, which increases with the PMMA deposition thickness (Fig. S13b). The cross-sectional views of the PMMA-coated membrane presented in Fig. S7 show that the PMMA molecules coat the sidewalls of the holes to some degree, but do not fill them completely, supporting our experimental findings.

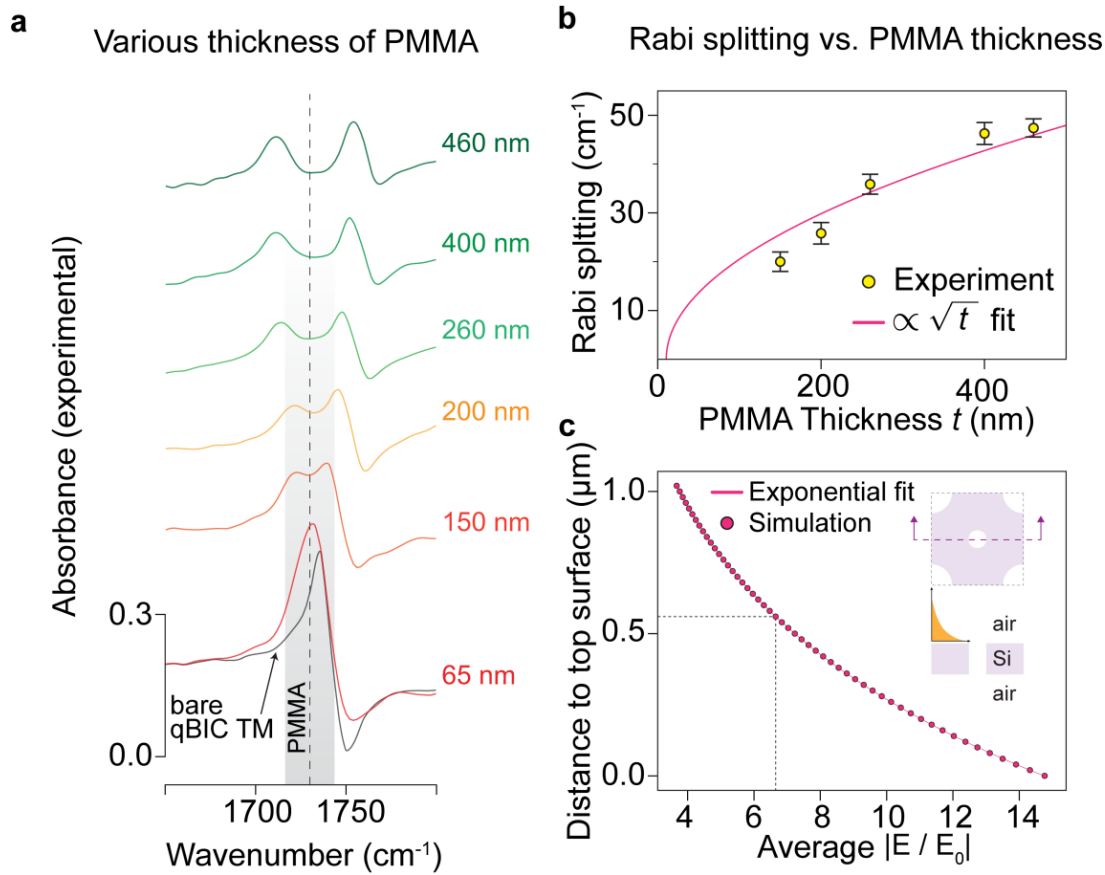

**Fig. 6: VSC strength as a function of molecule quantity in the photonic cavity. a,** Spectral response of VSC between the qBIC TM mode and PMMA C=O vibrational band (shaded gray) as the PMMA thickness filling the cavity increases. The bare metasurface resonance is also shown in black. Vertical offsets were applied to improve clarity. **b,** Experimentally measured (yellow dots) and analytical fit $\propto \sqrt{t}$ of Rabi splitting values as a function of PMMA thickness $t$. The Rabi splitting increases as the coupling strength is enhanced with a growing number of PMMA molecules that interact with the cavity. **c,** Field decay profile of the qBIC-TM mode on top of the membrane as a function of distance along the z-axis going away from the surface.

### Discussions

Judging from the progress in the past decade, chemistry under VSC is poised to be a new promising method for altering the properties of matter and controlling chemical processes[47]. Despite the progress, substantial knowledge gaps persist in our comprehension of the fundamental principles underpinning cavity-coupled chemistry. These include the roles of coherence and collective VSC on the modification of individual molecules' ground state properties[48], the effects of dark states in an ensemble coupled system, and VSC to electromagnetic vacuum fluctuations in the absence of external illumination[47]. Similarly, the potential technological impacts of polaritonic chemistry are extensive, offering the capability to steer chemical reactions toward desirable products[49] and to modify enzyme activity via VSC with water[50,51]. Realizing the full potential of polaritonic chemistry in the wider realm of biochemical sciences requires a cross-disciplinary strategy, particularly leveraging the diverse variety of optical resonators and investigation methods[23,52] developed by the photonics community. These can provide novel insights into the intricate mechanisms of VSC and facilitate the adoption of polaritonic chemistry on an industrial scale. In this context, Si-membrane metasurfaces compatible with CMOS manufacturing processes are poised to make significant contributions to this burgeoning field.

In order to advance the use of dielectric metasurfaces for VSC applications, our unique Si-membrane metasurface design introduces multi-fold benefits: 1) The optical resonance properties are not perturbed by the substrates, which have higher refractive index than air, and decrease the effective refractive index contrast between the resonator and its surrounding media (Fig. S11). 2) The hotspots surrounding the metasurface are formed in the air, thus generating accessible photonic cavities. 3) IR-transparent substrates are usually not compatible with high-throughput wafer-scale manufacturing processes. Our free-standing membrane design is fully compatible with CMOS manufacturing as they can be fabricated using conventional silicon-on-insulator wafers. 4) Our silicon-based metasurface will be compatible with existing microfluidics technology, making them compatible with liquid samples. 5) Vacuum photonic cavities that punch through a thin and robust membrane are well suited for applications requiring dynamic light-matter interactions where the liquid or air samples can flow through the holes while optical data is collected in real-time. 6) They have competitive mode volumes below 1 $(\lambda/n)^3$ per unit cell, even compared to the commonly used Fabry-Perot cavity or whispering gallery mode[41]—mode volumes 5 $(\lambda/n)^3$ and 6 $(\lambda/n)^3$, respectively. 7) As has been highlighted recently[25], cavity-molecule loss matching to achieve critical coupling strengthens the light-matter interactions. Since our metasurface design allows for resonance frequency and loss rate tunability, it is well suited for VSC applications. Furthermore, in this work, the metasurface cavity loss that we used for strong coupling experiments was intentionally kept high due to spectral resolution limitation of our measurement instrument. In future work, we plan to use a finely tunable illumination source and metasurfaces designed with sharper resonances to demonstrate higher Rabi splitting values.

Our experimentally acquired Rabi splitting values were not as high as the simulation-predicted results (Fig. 5a, b, and c vs Fig. 5d and S8), which might originate from the two following reasons. Firstly, unlike the simulation model with a uniform and continuous layer of PMMA coating, in real-world measurements, the PMMA layer is not a continuous film (see SEM images in Fig. S7) due to the presence of the metaunit. Moreover, the PMMA layer thicknesses were not directly measured on the metasurface, but on an unpatterned Si substrate using ellipsometry and reflectometry methods. The presence of flow-through holes may decrease the PMMA layer thickness, and result in a lower number of cavity-coupled molecules ($N$), and a lower Rabi splitting. Secondly, fabricated metasurfaces are not perfect geometries as in the simulation models with sharp edges, and it is expected the resonance mode to be stronger in the simulation than in the actual devices with inherent fabrication imperfections.

To conclude, in this work, we fabricated and characterized free-standing Si-membrane metasurfaces supporting qBIC resonances for VSC studies. We leveraged the Brillouin zone folding method to generate robust qBIC resonances that can efficiently trap mid-IR radiation in air cavities. Optically characterizing the fabricated metasurfaces, we identified the supported resonance types, which concurred with our simulation results. We measured a very high Q-factor at around 722, which is very competitive compared to previously reported mid-IR metasurface-based resonators (Table S1). Finally, we showcased the VSC between the qBIC resonances and PMMA's C=O vibrational resonance at 1730 cm$^{-1}$ by measuring the signature Rabi mode splitting over various detuning frequencies $\delta = \omega_{PMMA} - \omega_{qBIC}$. Our experimental findings closely matched the numerically calculated spectral anti-crossing pattern. Moreover, we measured VSC with various PMMA thicknesses and showed the $\sim\sqrt{N}$-correlation between VSC strength and the number of molecules, *N*, coupled to a cavity. Altogether, we introduced a new class of free-standing membrane dielectric metasurfaces that are completely CMOS manufacturing compatible and can impact polaritonic light-matter interactions by enabling VSC in powerful open cavities in the mid-IR.

**Methods**

**Numerical simulations**
The transmission spectra of the metasurfaces, except the ones in Fig. S2 and S11, were calculated using a finite-element frequency-domain solver (CST Microwave Studio 2023, Dassault Systèmes, France). The band diagrams in Fig. 3 and Fig. S2, and the transmission spectra in Fig. S2 and Fig. S11 were calculated using Tidy3D (Flexcompute, California, USA), a commercially available, large-scale finite-difference time-domain (FDTD) solver[53]. The metasurfaces were designed based on the simulations using periodic boundary conditions along the x- and y-directions and perfectly matching layer (PML) along the z-direction. To tune the TM qBIC resonances to the absorption band of the PMMA C=O bond (1730 cm$^{-1}$), we used the following geometric parameters for the double-hole design: period *P*=3.54 µm, large hole radius $r_1$=0.9 µm, small hole radius $r_2$=0.54 µm and membrane thickness of 1 µm. All structures were illuminated using a Gaussian beam with adequate spectral coverage from the top of the structure at normal incidence. For band diagram calculations, field time monitors were placed inside the membrane; for transmission spectra calculations, field monitor was placed underneath the metasurface. The cavity mode volumes were calculated using the Eqn. 3 from[54]:

$$V_{eff} = \frac{\int \varepsilon(r)|E(r)|^2 d^3r}{\max[\varepsilon(r)|E(r)|^2]} \quad (3)$$

**Fabrication of free-standing membrane dielectric metasurfaces**
Single-crystal 1-µm thick intrinsic Si-membranes (2.8 mm × 2.8 mm), supported by a 300 µm Si frame (10 mm × 10 mm), were used as substrates to fabricate the metasurfaces (Norcada, Alberta, Canada). Intrinsic Si was chosen to minimize loss and thus maximize Q-factor — the effect of doping can be seen in Fig S10. The membranes were quite robust with Young's modulus approximated to be similar to a 2-µm thick polycrystalline Si membrane at 140 GPa[55]. That is similar to bulk Si with Young's modulus at 160 GPa[56], and mirrors our fabrication experience with high yield at about 95%. Most of our membranes were fabricated at University of Wisconsin-

Madison Nanoscale Fabrication Center facility, and the fabrications were done as follows: To pattern the metasurface designs, a polymethyl methacrylate (PMMA-A8, Kayaku Advanced Material, Massachusetts, USA) resist layer was spin-coated at 4000 rpm and baked for 10 min at 180°C. Metasurface areas of 300 μm × 300 μm were then patterned by electron beam lithography using a 100 keV electron beam. After exposure, the PMMA was developed in MIBK: IPA 1:3 solution for 90 s. Subsequently, the holes in the membrane were formed using deep reactive ion etching (RIE), and the PMMA was stripped by acetone. To remove any PMMA residues from the dielectric metasurfaces, a 6-minute oxygen plasma treatment step was performed.

Due to infrastructure issues interrupting our access to the University of Wisconsin-Madison's cleanroom, some of the membranes were fabricated at the Argonne National Laboratory's Center for Nanoscale Materials as follows: metasurfaces were fabricated using a process sequence involving electron beam lithography, lift-off, and reactive ion silicon etching. Initially, a ~50 nm thin layer of $SiO_2$ hard mask was deposited on the bare Si-membrane using Plasma-Enhanced Chemical Vapor Deposition (PECVD) at 100°C, 2.5 mT, 1200 W inductive coupling plasma (ICP), 20 sccm $N_2O$, and 8.5 sccm $SiH_4$. This was followed by spin-coating ~110 nm of ZEP resist (1:1) at 4000 rpm and baking it for 3 minutes at 160°C. Metasurfaces of 300 × 300 μm² were then patterned using a 100 keV electron beam (4 nA and a dose of 250 μC/cm²). After the resist exposure, the ZEP on the Si membranes was developed in n-Amyl acetate for 60 seconds, followed by soaking in IPA for 60 seconds. Next, the hole shapes were imprinted in the thin $SiO_2$ layer using reactive ion $SiO_2$ etching at 20°C, 10 mT, 50 W RIE, 50 sccm $CHF_3$, 2 sccm $O_2$, and 10 T He backside pressure. The residual ZEP was stripped off using oxygen plasma (170 mT, 24 sccm $O_2$, 150 W for 3 minutes). The geometry was transferred from the $SiO_2$ into the 1 μm-thick Si-membrane using reactive ion silicon etching with HBr (20°C, 10 mT, 20 sccm $Cl_2$, 300 W RIE, 1000 W ICP, 10 T He backside pressure) for 6 seconds and then (20°C, 12 mT, 50 sccm HBr, 2 sccm $O_2$, 100 W RIE, 250 W ICP, 10 T He backside pressure) for 8 minutes. Finally, the residual $SiO_2$ was removed using reactive ion $SiO_2$ etching. No perceptible difference in both structural and optical response was found between metasurfaces fabricated in different facilities.

**PMMA spin coating and thickness characterization**
Various thicknesses of PMMA were obtained by combining the usage of PMMA (Kayaku Advanced Material, Massachusetts, USA) with different molecular weights, dilutions in anisole, and spin-coating speeds. Afterwards, the membrane was baked at 180°C for 5 minutes. All PMMA spin-coating was done only from the top side of the metasurface. To measure the PMMA thickness, for each metasurface coated with PMMA, a bare Si chip of the same size was also PMMA coated using identical parameters. We used an ellipsometer (J.A. Woollam, Nebraska, USA) and a reflectometer (Filmetrics, California, USA) to measure PMMA thickness on bare Si substrates.

**Optical setup and measurements**

Mid-IR spectral measurements were done, unless otherwise noted, using a tunable quantum cascade laser (QCL) integrated into a mid-IR microscope (Spero-QT, Daylight Solutions, California, USA). Using four QCL modules, the microscope can collect spectra covering the fingerprint spectral region from 950 to 1800 cm$^{-1}$ with 2 cm$^{-1}$ spectral resolution. During acquisition, the sample chamber was continuously purged with dry nitrogen to clear out water vapor. The spectral data was acquired in a transmission mode, where the sample was illuminated using collimated light at normal incidence, and the transmittance was collected using a 12.5 × IR objective (pixel size 1.3 µm, 0.7 NA) and detected using an uncooled microbolometer focal plane array with 480 × 480 pixels obtaining a field of view of 650 µm × 650 µm. In transmission mode, our chemical microscope captures the sample's decadic absorbance as $A$ = -log$_{10}$($T$), where the transmittance $T = I/I_0$ and $I_0$ is the transmission through unpatterned Si-membrane as a sample. From the definition of decadic absorbance, we then calculate the transmittance of the sample as follows: $T = 10^{-A}$ and, therefore, $1-T = 1- 10^{-A}$. Some previous works[57,58] calculated absorbance through the subtraction of transmittance and reflectance from incident light, i.e., $A=1-R-T$ to quantify Rabi splitting. However, in our work, the absorbance is derived only from transmittance following the common practice often found in literature[49,59], and it yields similar results to $A=1-R-T$ based on our simulation-based investigation (Fig. S19).

Fig. 3d, 3e, S5, and S12 were composed using data collected by a Fourier-transform infrared (FTIR) spectrometer coupled to an infrared microscope (Bruker Vertex 70 FTIR and Hyperion 2000). Transmittance spectra were collected with unpolarized light through a refractive objective (5×, 0.17 NA, Pike Technology, Wisconsin, USA) and measured by a liquid-nitrogen-cooled mercury-cadmium-tellurium (MCT) detector. The condenser lens was omitted for normal incidence illumination, see Fig. S12 for its effects on the measured Q-factor. All acquired transmittance spectra of the Si-membrane metasurfaces using both FTIR and QCL microscope were normalized using the transmission spectrum of an unpatterned Si-membrane for source normalization.

**Data processing**
The acquired data was processed by custom-made scripts using MATLAB (MathWorks, Massachusetts, USA). For baseline correction, a third-degree polynomial was fitted to each spectrum collected from metasurfaces and then subtracted from the signal to remove the Fabry-Perot background caused by the 1-µm thick Si-membrane. For the VSC measurements shown in Fig. 5, 6, S6 and S9 image registrations were performed between the hyperspectral image datasets collected before and after PMMA coating to ensure a comparison of the same metasurface regions. For presentation purposes, the data in Fig. 4, and Fig. 5a were further normalized so that their highest value is set to one.

To determine polariton spectral peak positions, the second derivative of the spectrum was calculated and smoothed using a Savitzky–Golay (SG) filter with 2$^{nd}$ polynomial order and a window width of 13 data points. The dips in the second derivative reveal the spectral positions of the peaks in the spectrum, and those of upper and lower polaritons were identified to obtain the Rabi splitting Ω. The error bars in Fig. 5b, 5c, and 6b comes from the standard deviation of the 11x11 hyperspectral pixels, i.e., 121 spectra used for the determination of the bare peak positions and Rabi splitting distances. The detuning in Fig. 5b was determined as the frequency with which the cavity achieves the lowest Rabi splitting (shown in Fig. 5b). Detuning distance is then calculated through the difference between the wavenumber of bare cavity peak that achieves zero detuning upon coating with PMMA to the wavenumber of bare cavity peak of the other cavity of interest. The analytical fit in figure 6b was fitted to the equation $\Omega = \sqrt{4g^2 - (\gamma_{PMMA} - \gamma_{qBIC})^2}$, where $4g^2$ was assumed to be the PMMA thickness multiplied by a factor that is then determined through the fit, and the values of 13.5 cm$^{-1}$ and 6.5 cm$^{-1}$ were used for $\gamma_{PMMA}$ and $\gamma_{BIC}$, respectively. Moreover, the signals of PMMA of corresponding thickness on unpatterned metasurface were subtracted from the signals with thickness ≥ 260 nm in Fig. 6a to account for weakly interacting PMMA, and thus increase the visibility of the polariton formations. Their raw data are shown in Fig. S18. Finally, all cavity losses and Q-factors were obtained by fitting the transmittance $T$ with the following Fano fit[60] (Eqn. 4).

$$T = \left| ie^{i\phi} t_0 + \frac{\gamma_r}{\gamma_r + \gamma_a + i(\lambda - \lambda_{res})} \right|^2 \quad (4)$$

Where $ie^{i\phi} t_0$ describes the background and the shape of the resonance. The Q-factor can then be calculated as in Eqn. 5.

$$Q = \frac{\lambda_{res}}{2(\gamma_r + \gamma_a)} \quad (5)$$

Where $\lambda_{res}$, $\gamma_r$, and $\gamma_a$ are the resonance wavelength, radiative, and absorptive loss rates, respectively.

**Data availability**

The data generated for this study has been provided in the main text and supplementary information.

**Acknowledgement**

F.Y. acknowledges financial support from the National Institutes of Health (grant no. R21EB034411) and the National Science Foundation (grant no. 2401616). The authors thank Prof. Eduardo R. Arvelo from the Electrical & Computer Engineering Department at UW-Madison for assistance in illustration. The authors thank Prof. Mikhail Kats for providing access to the FTIR. The authors thank Prof. Yuri Kivshar from Australian National University for fruitful discussions. Work performed at the Center for Nanoscale Materials, a U.S. Department of Energy Office of Science User Facility, was supported by the U.S. DOE, Office of Basic Energy Sciences, under Contract No. DE-AC02-06CH11357.

**Authors Contribution**

W.A. characterized and simulated the metasurface, processed the data, and drafted the paper. S.R. fabricated the metasurface, helped with the metasurface characterization, and drafted the paper. A.B. and S.K.B. helped with the simulation of the metasurface. H.M. helped with the characterization of the metasurface. D.A.C. helped with the fabrication of the metasurface. F.Y. conceived the idea, designed the experiments, supervised the project, and drafted the paper. All authors revised the paper.

**Competing Interest**

The authors declared no competing interest.